\documentclass[a4paper,fleqn,usenatbib]{mnras}

\usepackage{newtxtext,newtxmath}

\usepackage[T1]{fontenc}
\usepackage{ae,aecompl}

\usepackage{graphicx}	
\usepackage{amsmath}	
\usepackage{amssymb}	

\title[A spectroscopic search for asymmetries in WR137]{An extensive spectroscopic time series of three Wolf-Rayet stars  $-$  II. A search for wind asymmetries in the dust-forming WC7 binary WR137}

\author[N. St-Louis et al.]{N. St-Louis,$^{1}$\thanks{E-mail: stlouis@astro.umontreal.ca},
C. Piaulet$^{1}$,
N. D. Richardson$^{2}$,
T. Shenar$^{3}$,
A.F.J. Moffat$^{1}$,
\newauthor
T. Eversberg$^{4,5,6}$,
G.M. Hill$^{7}$,
B. Gauza$^{8,9}$,
\newauthor
J.H. Knapen$^{10,11}$,
J. Kub$\acute{\rm a}$t$^{12}$,
B. Kub$\acute{\rm a}$tov$\acute{\rm a}$$^{12}$,
D.P. Sablowski$^{4,5,13}$,
\newauthor
S. Sim\'on-D\'\i az$^{10,11}$,
F. Bolduan$^{5}$,
F.M. Dias$^{4,5}$,
P. Dubreuil$^{5,14}$,
D. Fuchs$^{5}$,
\newauthor
T. Garrel$^{5,14}$,
G. Grutzeck$^{4,5}$,
T. Hunger$^{4,5}$,
D. K\" usters$^{4,5}$,
M. Langenbrink$^{5}$,
\newauthor
R. Leadbeater$^{4,5,15}$,
D. Li$^{4,5,16}$,
A. Lopez$^{5,14}$,
B. Mauclaire$^{5,14}$,
T. Moldenhawer$^{5}$,
\newauthor
M. Potter$^{5,17}$,
E.M. dos Santos$^{5}$,
L. Schanne$^{4,5}$,
J. Schmidt$^{5}$,
H. Sieske$^{5}$,
\newauthor
J. Strachan$^{5,18}$,
E. Stinner$^{5}$,
P. Stinner$^{4,5}$,
B. Stober$^{4,5}$,
K. Strandbaek$^{4,5}$,
\newauthor
T. Syder$^{5}$,
D. Verilhac$^{5,14}$,
U. Waldschl\" ager$^{4,5}$,
D. Weiss$^{4,5}$ 
and A. Wendt$^{5}$
\\
$^{1}$D\'epartement de Physique, Universit\'e de Montr\'eal, C.P. 6128 Succ. Centre-ville, Montr\'eal, QC H3C 3J7, Canada\\
$^{2}$Department of Physics and Astronomy, Embry-Riddle Aeronautical University, 3700 Willow Creek Road, Prescott, Arizona 86301, USA\\
$^{3}$Instituut voor Sterrenkunde, Celestijnenlaan 200D bus 2401, 3001 Leuven, Belgium\\
$^{4}$VdS Section Spectroscopy, Germany\\
$^{5}$Teide Pro-Am Collaboration\\
$^{6}$Schn\"orringen Telescope Science Institute, Ringweg 8a, 51545 Waldbr\"ol, Germany\\
$^{7}$W. M. Keck Observatory, 65-1120 Mamalahoa Highway, Kamuela, HI 96743, USA\\
$^{8}$Centre for Astrophysics Research, School of Physics, Astronomy and Mathematics, University of Hertfordshire, College Lane, Hatfield AL10 9AB, UK\\
$^{9}$Janusz Gil Institute of Astronomy, University of Zielona G\'{o}ra, Lubuska 2, 65-265 Zielona G\'{o}ra, Poland\\
$^{10}$Instituto de Astrof\'\i sica de Canarias, E-38200 La Laguna, Tenerife, Spain\\
$^{11}$Departamento de Astrof\'\i sica, Universidad de La Laguna, E-38206 La Laguna, Spain\\
$^{12}$Astronomick\'y \'ustav, Akademie v$\check{e}{\rm d}$ $\check{C}{\rm esk\'e}$ Republiky, 251 65 Ond\v{r}ejov, Czech Republic\\
$^{13}$Leibniz-Institut for Astrophysics Potsdam (AIP), An der Sternwarte 16, D-14482 Potsdam, Germany\\
$^{14}$Astronomical Ring for Access to Spectroscopy (ARAS), France\\
$^{15}$Three Hills Observatory, The Birches, Torpenhow, Wigton CA7 1JF, UK \\
$^{16}$Jade Observatory, Jin Jiang Nan Li, Jin Jiang Road, He Bei District, 501-47-59 Tianjin, China\\
$^{17}$Beverly Hills Observatory, PO Box 3626, Baltimore, MD 21214, USA\\
$^{18}$School of Physics and Astronomy, Queen Mary University of London, 327 Mile End Rd., London E1 4NS, UK\\
}

\date{Accepted XXX. Received YYY; in original form ZZZ}

\pubyear{2020}

\begin{document}
\label{firstpage}
\pagerange{\pageref{firstpage}--\pageref{lastpage}}
\maketitle

\begin{abstract}
We present the results of a four-month, spectroscopic campaign of the Wolf-Rayet dust-making binary, WR137.  We detect only small-amplitude, random variability in the C{\sc iii}$\lambda$5696 emission line and its integrated quantities (radial velocity, equivalent width, skewness, kurtosis) that can be explained by stochastic clumps in the wind of the WC star.  We find no evidence of large-scale, periodic variations often associated with Corotating Interaction Regions that could have explained the observed intrinsic continuum polarization of this star. ÕOur moderately high-resolution and high signal-to-noise average Keck spectrum shows narrow double-peak emission profiles in the H$\alpha$, H$\beta$, H$\gamma$, He{\sc ii}$\lambda$6678 and He{\sc ii}$\lambda$5876 lines.  These peaks have a stable blue-to-red intensity ratio with a mean of 0.997 and a root-mean-square of 0.004, commensurate with the noise level; no variability is found during the entire observing period. We suggest that these profiles arise in a decretion disk around the O9 companion, which is thus an O9e star.  The characteristics of the profiles are compatible with those of other Be/Oe stars. The presence of this disk can explain the constant component of the continuum polarization of this system, for which the angle is perpendicular to the plane of the orbit, implying that the rotation axis of the O9e star is aligned with that of the orbit. It remains to be explained why the disk is so stable within the strong ultraviolet radiation field of the O star. We present a binary evolutionary scenario that is compatible with the current stellar and system parameters.\end{abstract}




\section{Introduction}
\subsection{The long-Period Binary WR137}

The Wolf-Rayet (WR) star WR137 (=HD192641), of spectral type WC7, was first suspected to be a long-period binary by \citet{1985MNRAS.215P..23W} who found a slow increase of its 3.6 $\mu m$ infrared (IR) light. They associated the rising flux to an increase in the stellar mass-loss rate and to condensation of dust, most likely in a shock cone formed as a consequence of colliding winds in a massive binary.  \cite{1981ApJ...246..145M} found that the star did not show radial velocity variations on a short timescale and therefore that is was not a close binary system. This was later confirmed by \citet{1986AJ.....91.1392M}. However, the presence of absorption lines superposed on its broad emission lines, that were also known to be diluted, hinted to a possible long-period system.
 
The long-period binary nature of this star was further supported by \citet{1991IAUS..143..245A} who found radial velocity variations of absorption and emission lines in anti-phase with a period of $\geq$ 4400 days. A few years later, \citet{1995IAUS..163..231A} confirmed that the star was indeed a binary by presenting the first orbit of this system. A period of 5680 days and an eccentricity of 0.07 were determined but the data were noisy and the radial velocity amplitudes of both stars were found to be quite low: 30.5$\pm$1.1 km s$^{-1}$ for the WR star and 21.6$\pm$3.7 km s$^{-1}$ for the absorption-line star.  The period was once again revised some years later by \citet{2001MNRAS.324..156W} who presented the first IR light-curve covering one complete orbit. The period was found to be P=4765$\pm$50 days or 13.05$\pm$0.15 years. The dust generating the IR light-excess was imaged using the Hubble Space Telescope by \citet{1999ApJ...522..433M} who estimated the total dust mass to be around 0.1 M$_{\oplus}$.  By assuming that the trajectory of this dust was in the orbital plane, these authors deduced a high orbital inclination of 68$^{\rm o}$. 

The system was revisited with higher quality spectroscopic data by \citet{2005MNRAS.360..141L} who presented the first orbit of the system with a period in agreement with that determined from the IR light-curve (P=4766$\pm$66 days). Adopting an O9 spectral type for the companion \citep{2001NewAR..45..135V} with a luminosity class between III and V and therefore  a mass of M$_{\rm O}$=20$\pm$2 M$_{\odot}$, they concluded that the implied inclination was 67$^{\rm o}$ and the mass of the WR star was M$_{\rm WR}$=4.4$\pm $1.5 M$_{\odot}$, which is on the low side for a WR star.

Finally, the H-band interferometric data of \citet{2016MNRAS.461.4115R} allowed them to separate the two components of the binary.  At that wavelength, they find that the WR star contributes the greater part of the total flux the combined binary flux (f$_{\rm WR}=0.59 \pm0.04$) and their data support a high, nearly edge-on inclination. They also  present spectroscopic modelling of both components of the binary (using the Keck mean spectrum obtained for this campaign), with the PoWR code \citep{2004A&A...427..697H}, which allowed them to determine the fundamental parameters of both stars.  They find a spectral type of WC7pd+O9V.

\subsection{A Non-Spherical Stellar Wind}

WR137 has long been know to be linearly polarized. \citet{1951ApJ...114..241H} measured an optical broadband polarization, presumably mostly interstellar (IS) in origin, of P=1.2\%\ and a polarization angle of 168$^{\rm o}$. \citet{1989ApJ...347.1034R} found that the broadband polarimetric flux of this star was only slightly variable with an average value of P=1.175\%\ , a standard deviation of 0.020\%\ and a typical error bar of 0.011\%\  hinting that some small fraction of this polarization was intrinsic to the star. This was later confirmed by \citet{1998MNRAS.296.1072H} \citep[see also ][]{1994Ap&SS.221..365H} who found reduced polarization at emission-line wavelengths -- the so-called "line effect" -- generally interpreted as caused by the dilution of polarized continuum flux by unpolarized (or less polarized) line emission. From the difference between the continuum and line polarization they estimate the intrinsic continuum polarization for WR137 to be P$_c$=0.57$\pm$0.20 \%.   Because of the large binary separation at the time of their observations,  they concluded that it was unlikely that the polarization was caused by the "binary effect" (i.e. the fact that the O star is an asymmetric light source from the point of view of the scattering electrons in the WR wind) and attributed it instead to an asymmetry intrinsic to the WR wind, more specifically to a flattening of the wind by rapid rotation. Using the expression for the binary-induced amplitude of polarization variability from \citet{1988ApJ...330..286S} with i=68$^{\rm o}$ \citep{1999ApJ...522..433M}, e=0.178 and a binary separation  ranging from 2290 to 3370 R$_{\odot}$ \citep{2005MNRAS.360..141L}, v$_{\infty}$=1885 km/s and $\dot { \rm M}$=3$\times$10$^{-5}$ M$_{\odot}$/yr \citep{1990ApJ...361..607P} and f$_c$=0.86 \citep{2016MNRAS.461.4115R}, we estimate the expected polarization variability from the binary to be $\sim$0.02\%, which is of the same order of magnitude as the short-timescale, non-periodic polarization variability measured by \citet{1989ApJ...347.1034R}.  We therefore concur that the intrinsic polarization level of WR137 measured by \citet{1998MNRAS.296.1072H} is unlikely to be the result of the binary effect.

  The most recent linear polarimetric study of WR137 was presented by \citet{2000A&A...361..273H} who obtained multi-epoch spectropolarimetry of WR137. They confirmed that the continuum polarization is indeed variable but that the polarization angle is remarkably constant. After carefully estimating the interstellar polarization by fitting a standard Serkowski law, finding q$_{IS}$=0.80$\pm$0.1\%, u$_{IS}$=$-$0.64$\pm$0.1\%, they subtracted it vectorially from the observed polarization and found that the intrinsic polarization angle is $\sim$17$^{\rm o}$, i.e. nearly perpendicular to the extended dust emission (see their Figure 4) observed by  \citet{1999ApJ...522..433M} and thought to arise in the colliding-wind shock cone. This indicates that the asymmetry found in the WR wind is mirrored in the shock cone, which is located near the O star far from the WR component.  They conclude that the WR component of WR137 has a flattened wind with an equator-to-pole density ratio of between two and three and that this geometry is stable over long periods of time. They attribute the small-scale random variability in the level of continuum polarization to the presence of inhomogeneities in the WR wind. Finally, the authors suggest that WR137 presents striking similarities with two other WR stars that show line depolarization, WR6 and WR134, which have been interpreted as harbouring in their wind large-scale Corotating Interaction Regions (CIRs). However, those two stars have also been shown to present periodic line-profile variations in the optical \citep[eg.][]{1995ApJ...452L..57S,1999ApJ...518..428M} that can be attributed to such structures but no intense spectroscopic monitoring campaign over an extended period of time exists in the literature for WR137. \citet{2005MNRAS.360..141L} did find a period of 0.83 days in the absorption troughs of the C{{\sc iv}}$\lambda$5802/12 and He{\sc {i}}$\lambda$5876 lines but not in the emission components, as are thought to arise from CIRs in WR winds.  Furthermore, the variability in the absorption components did not show the usual characteristics of Discrete Absorption Components (DACs) thought to originate from CIRs, as seen in the ultraviolet for O stars. We note that \citet{2012A&A...547A..83G} compiled all the spectropolarimetric information on Galactic WR stars. In addition to WR6, WR134 and WR137 they find that only WR16, WR40 and WR136 show line depolarization.  We might add that the WN4b star WR1, also exhibits depolarized stellar lines \citep{2013ApJ...777....9S} and has been shown to present periodic optical line-profile variations \citep{2010ApJ...716..929C}.

In this paper we present the results of an intense optical spectroscopic monitoring campaign of WR137. In 2013, professional and amateur astronomers joined forces to carry out a four-month long worldwide observation run of three bright WR stars, WR134, WR135 and WR137, the original three WR stars discovered by \citet{1867CRAS...65..292W}. Results for the WN6 star WR134 were presented in the first of this series of papers by \citet{2016MNRAS.460.3407A}.  Section 2 summarizes the spectroscopy of WR137 collected from 7 different sites across the world. In Section 3, we present the measurements made for the isolated, strong C{\sc iii}$\lambda$5696 line and for the H$\alpha +$He{\sc ii} complex near 6560\AA. Section 4 presents a discussion of the nature of this binary system based on the findings of our campaign dataset. We conclude in Section 5.

\section{Observations and Data Reduction}

The data for WR137 within the framework of our 2013 campaign were collected between 26 May and 1 October 2013 using seven different telescopes. Details of these observations are listed in Table\,\ref{OBS}. Our amateur astronomer co-authors ("The Teide Pro-Am collaboration"; see affiliation 5 in the authorship list) collected the majority of the data at the Teide Observatory of the Instituto de Astrof$\acute{\i}$sica de Canarias (IAC) in Tenerife using the 0.82m IAC80 telescope and a fibre-fed echelle spectrograph (eShel) manufactured by Shelyak Instruments (France)\footnote{http://www.shelyak.com/?lang=en} combined with a CCD. The spectrograph and CCD were kindly supplied by B. Stober.   Two of us obtained spectra using private instrumentation (Robin Leadbeater, RL, and Mike Potter, MP). Unfortunately, the data obtained by Mike Potter were subsequently found to be corrupted by mercury emission lines from city streetlights and were not used in our analysis.  Despite the small telescope sizes, the data are of high quality in both signal-to-noise (S/N) and resolution.  Observations were also collected at four professional observatories:  (1) the Dominion Astrophysical Observatory (DAO) in British Columbia, Canada, (2) the Ond$\check{\rm r}$ejov Observatory at the Astronomical Institute of The Czech Academy of Sciences in the Czech Republic, (3) the Observatoire du Mont-M\'egantic (OMM) located in Qu\'ebec, Canada and (4) the Keck Observatory located in Hawaii, USA.  For all these observing settings, calibration frames necessary for data reduction were obtained each night (bias, dark and flat field frames) and later used in the data calibration. In addition, ThAr emission spectra from standard discharge lamps were secured before or after each exposure for the purpose of wavelength calibration.   

\begin{table*}
	\centering
	\caption{List of observatories and individual observers that contributed to the campaign. We list the telescope, spectrograph and CCD used, the Heliocentric Julian Date (HJD) range, the total number of spectra and nights over which they were obtained, the resolving power, the wavelength range and the average S/N ratio. Note that there are two rows for OMM since there were two separate runs using different gratings.  The information given for the Keck data only corresponds to the 4$^{th}$ order in which the C{\sc iii}$\lambda$5696 line is located.  The S/N provided for the Keck spectrum is that of the average of all 28 spectra.}
\label{OBS}
	\begin{tabular}{lccccccccr} 
		\hline
		Observatory & Telescope & Spectrograph & CCD & HJD & N$_{spec}$&N$_ {nights}$& Resolving & $\lambda _{\rm coverage}$ & S/N \\
		&&&&--2450000&&&Power (${\lambda \over \Delta \lambda}$)&($\AA$)&\\
		\hline
		\multicolumn{9}{c}{\bf Professional Facilities}\\
		OMM (1) & 1.6m & Perkin-Elmer & STA0520 Bleu & 6491--6502 & 13 & 5&5300 &4500$-$6500 & 200 \\
		OMM (2) &  &  &  & 6564--6567 & 17 & 4&7000 & 4750$-$6100& 200 \\
		DAO  & 1.8m & Cassegrain & SITe-2 & 6474--6482 & 49 & 10&5500 & 5140$-$5980 & 200 \\
		Ond\v{r}ejov  & Perek (2m) & Coud\'e & PyLoN 2018$\times$512 BX & 6483--6542 & 18 & 8&10000 & 5500$-$6000 & 100 \\
		Keck  & Keck II (10m) & ESI & MIT-LL W62C2 & 6517 & 28 & 1& 13000 & 4000$-$10000 & 150 \\
		Teide  & IAC80 & eShel & Nova3200 & 6439--6552 & 165 & 62 & 10500 & 4500$-$7400 & 100 \\
		\multicolumn{9}{c}{\bf Individual Contributors}\\
		Leadbeater  & C11 (0.28m) & LHIRES III & ATIK-314L+ & 6486--6495 & 9 & 5& 5300 & 5580$-$5950 & 100 \\
		Potter  & C14 (0.36m) & LHIRES III & SBIG ST-8 & 6468--6543 & 9 & 4& 7500 & 5580$-$5950 & -- \\
		\hline
	\end{tabular}
\end{table*}

For this WC type star, we decided to prioritize the C{\sc iii}$\lambda$5696 emission line. The intensity of this line has been shown to be quite sensitive to wind parameters, particularly to wind density \citep{1989ApJ...347..392H, 1992A&A...255..200H, 2002A&A...392..653C}. This increases our chances of detecting any perturbations in the wind. This line is also relatively strong and well isolated. Therefore, all observers, amateurs and professionals, observed at least this spectral line. 

The IRAF\footnote{IRAF is distributed by the National Optical Astronomy Observatory, which is operated by the Associated Universities for Research in Astronomy, Inc., under cooperative agreement with the National Science Foundation.} software package was used to carry out the data reductions using standard techniques.  The blaze functions of the Teide and Keck echelle data were carefully determined and used to correct for the spectrograph response. This was necessary as the emission lines of the WR star are broad and may straddle two successive orders.  To do so, we made use of our observations of the standard A0V star, Zeta Aquilae.  For each night during which that star was observed, we first took an average of all available observations. We then fitted a cubic spline function to each order individually. Finally, we assembled all individual fits into one function. The result consists in our Blaze function for that particular night, which we applied to all spectra obtained during that same night. For the nights during for which we had no observations of Zeta Aquilae, we used the Blaze function from the closest night for which we had obtained an observation. Finally, the spectra were normalized in the vicinity of the C{\sc iii}$\lambda$5696 line and additionally in the region of the H$\alpha$ line for the Keck and Teide spectra.  We chose continuum regions which were as much as possible free of strong emission lines: 5525$-$5540~\AA , 5645$-$5650~\AA , 5747$-$5750~\AA\  and 5977$-$6050~\AA\ for the C{\sc iii} line and 6480$-$6512~\AA\ and 6630$-$6645~\AA\  for H$\alpha$.

\section{Data Analysis}

We show the complete spectroscopic dataset for C{\sc iii}$\lambda$5696 in Figure~1 using a different colour for each observatory. We also show the mean spectrum in black. We note that the C{\sc iii}$\lambda$5696 emission line is positioned on the eShel detector such that the part of the blue wing of the line suffers from low S/N in several spectra, as can be seen in the plot. However, the core of the line is relatively unaffected, allowing for a useful measurement of most quantities. We show in Figure~2 the Teide and Keck datasets in the region of the H$\alpha$ line, respectively in blue and red and the mean spectrum in black.  As can be readily seen in these figures, only small-scale variability is detected, most likely related to clumps in the wind of the WR star  \citep{1999ApJ...514..909L}. Contrary to WR134 \citep{2016MNRAS.460.3407A}, no large-scale changes that could be  related to the presence of CIRs in the wind of the WR component of WR137 are detected. 

Nevertheless, we have calculated spectroperiodograms using the Lomb-Scargle formalism, for the wavelength regions of the C{\sc iii}$\lambda$5696 line (5650-5750 \AA) and of the He{\sc ii}/C{\sc ii}/C{\sc iv}/H$\alpha$ complex (6495-6650 \AA). Apart from the 1-day aliasing peaks caused by the time sampling of the campaign, no significant periodicities were detected at any wavelength. The 0.8d periodicity in the more-sensitive P Cygni absorption components of some optical emission lines of WR137 as found by \citet{2005MNRAS.360..141L} could still be of relevance however, but remains to be confirmed as it was detected only at the 2\%\ level, similar to the random scatter. 
 
\begin{figure}
	\includegraphics[width=\columnwidth]{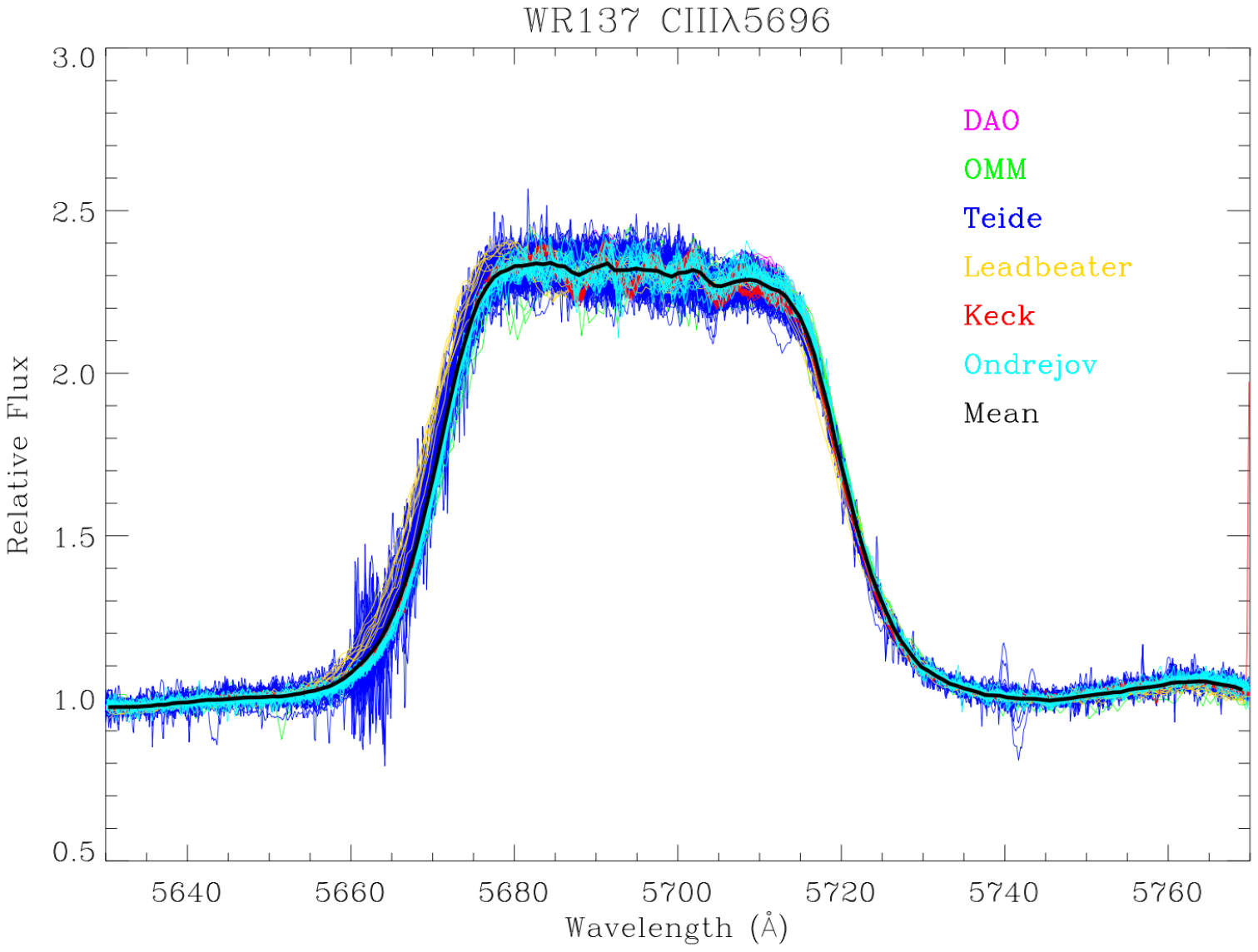}
    \caption{Superposition of all C{\sc iii}$\lambda$5696 line profiles in our dataset. We present observations from different observatories using different colours, as indicated in the labels. The black line is the average of all spectra.}
    \label{fig:fig1}
\end{figure}
\begin{figure}
	\includegraphics[width=\columnwidth]{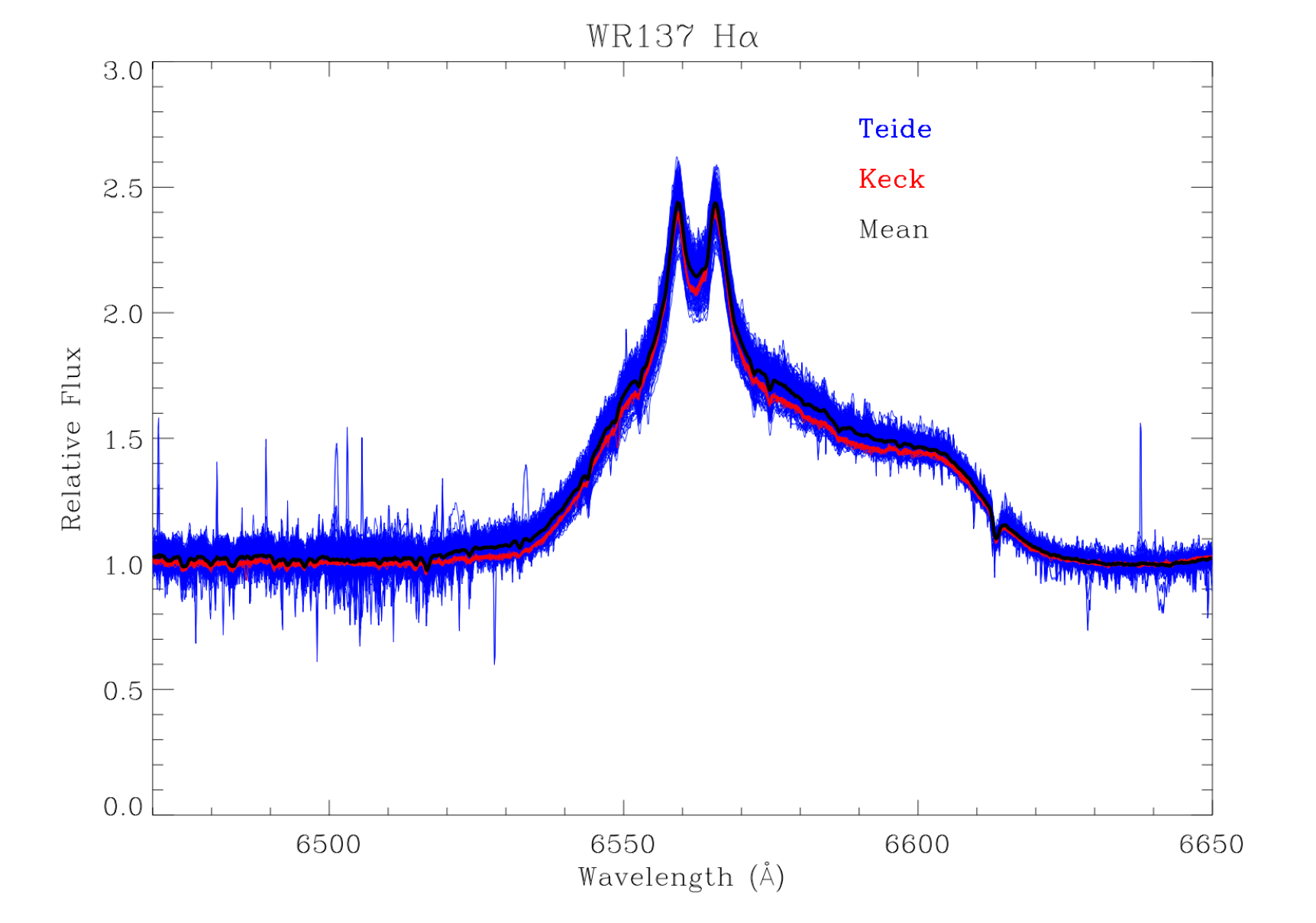}
    \caption{Superposition of all Keck and Teide observations in the wavelength region of the He{\sc{ii}}$\lambda$6560, C{\sc{ii}}$\lambda$6578 and C{\sc{iv}}$\lambda$6592 complex. The black line is the average of all spectra.}
    \label{fig:fig2}
\end{figure}

In order to describe the line variability as a function of time, we measured integrated quantities of the C{\sc iii}$\lambda$5696 profile.  All our measurements can be found in a table in text format available as online material. A sample of the content of this table can be found in Table\,\ref{moments}. We first calculated the radial velocities by using the average of the bisector of the line between two relative intensity levels (1.3 and 1.9) in order to avoid low fluxes as well as the peak of the line, which presents some small-scale variability. To remove difficult-to-avoid zero-point biases between the various observatories using different instrumentation, we removed the average radial velocity of each dataset.  The results are presented in the top panel of Figure~3 as a function of the Heliocentric Julian Date. Only small-scale variability is found with a standard deviation of 8.9 km/s, which is $2-3$ times the typical error bar of 3.4 km/s. This is in agreement with the results of \citet{1981ApJ...246..145M} who found no significant RV changes over a short time period. The RV changes from the long-period orbit \citep[K$_{WR}$=27.9$\pm$1.2 km/s][]{2005MNRAS.360..141L} would produce a very small gradual shift of the radial velocity of the star on the order of $\sim$5 km/s over the four-month observing period, well below our observed scatter. Note that these authors observed a similar short-timescale scatter (see their Figure 12). 

\begin{table*}
\centering
\caption{The various moments of the C{\sc iii}$\lambda$5696 emission line profile of WR137 from our dataset. }
\label{moments}
\begin{tabular}{ccccccccc}
\hline
    HJD-2450000 & EW (\AA ) & $\sigma_{EW}$ (\AA ) & Sk & $\sigma_{Sk}$ & Kr & $\sigma_{kr}$ & v (km/s) & $\sigma_v$ (km/s)\\
\hline
   6438.6201172 &  -71.1029816 &  0.5324892 &  0.0440514 &  0.0011094 &   2.1333027 &   0.0018192 & -10.4173193 &   3.1772523\\
   6440.5947266 &  -72.1759262 &  0.6781074 &  0.0655105 &  0.0014491 &   2.1079762 &   0.0023595 & -36.7564163 &  5.0564389\\
   6443.5424805 &  -70.6214600 &  0.6510331 &  0.0442086 &  0.0014122 &   2.1298163 &   0.0023164 &  -5.9160085 &  4.6156330\\
   6443.6972656 &  -70.5170593 &  0.5447067 &  0.0411231 &  0.0011515 &   2.1324668 &   0.0019046 &   1.9805768 &   4.4711733\\
   6445.5629883 &  -70.9593735 &  0.6368606 &  0.0291732 &  0.0013864 &   2.0568912 &   0.0022452 &  -7.4078712 &  4.5407028\\
...&&&&&&&&\\
\hline
\end{tabular}
\end{table*}

We also measured the total equivalent width of the line, which we present in the second panel from the top in Figure~3. Note that the initial measurements revealed systematic differences of the order of $2-3$\%\ between the spectra obtained with an echelle (Teide and Keck) and those obtained with a regular grating (OMM,Ond\v{r}ejov, DAO, Leadbeater, Potter). The echelle spectra were found to have systematically lower equivalent widths. We believe that the most likely cause for these differences is the removal of the blaze function in echelle spectra, which was fitted with a low order function and varies slowly across a spectral line. Another possible explanation would be an incomplete removal of the scattered light.  We chose the spectra obtained  at the DAO that show very little scatter (49 spectra over 8 consecutive nights) as a reference and applied a simple correction to the Teide and Keck fluxes using the following expression:

$$(F_{\rm ech, corr}-1)=\left(  {{\langle {\rm EW}_{\rm  DAO}\rangle \over \langle{\rm EW}_{\rm ech}\rangle }}\right)  (F_{\rm ech}-1),$$

and the equivalent widths with 
$${\rm EW}_{\rm ech,corr}=\left(  {\langle {\rm EW}_{\rm  DAO}\rangle \over \langle{\rm EW}_{\rm ech}\rangle }\right) {\rm EW}_{\rm ech},$$ which assumes that the line is roughly flat-topped. In these expressions, $F_{\rm ech, corr}$ and $F_{\rm ech}$ are the corrected and original fluxes of the echelle spectra respectively, $\langle$EW$_{\rm  DAO}\rangle $ and $\langle$EW$_{\rm ech}\rangle $ are the averages of the EW values of the C{\sc iii}$\lambda$5696 line obtained respectively at the DAO and for echelle data and EW$_{\rm ech,corr}$ and EW$_{\rm ech}$ are corrected and original values of the equivalent widths of individual line profiles.

\begin{figure*}
	\begin{center}\includegraphics[width=\textwidth]{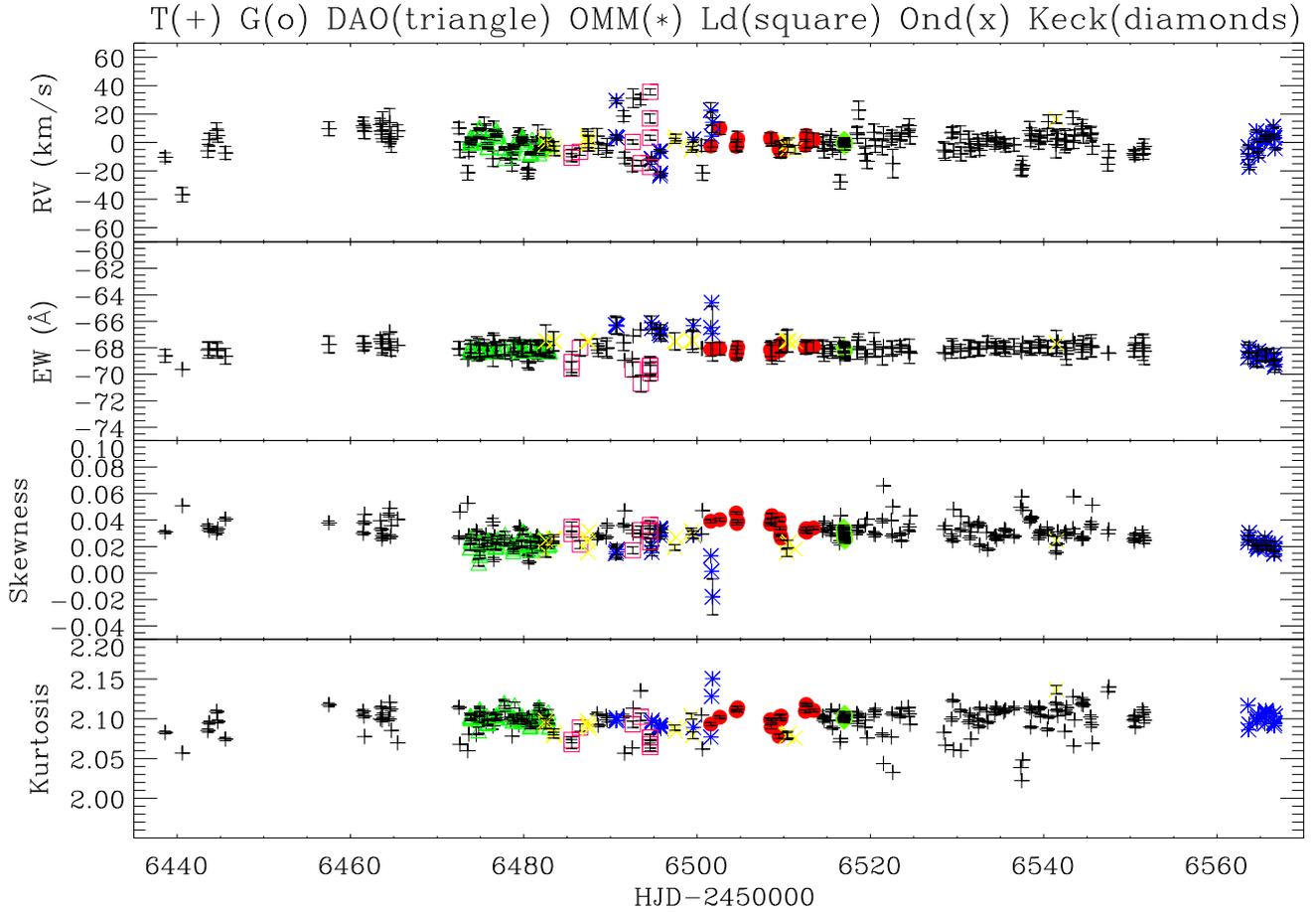}
    \caption{Radial velocities (RV), Equivalent Widths (EW) Skewness and Kurtosis of the C{\sc iii}$\lambda$5696 profile of WR137 obtained during our campaign. Different symbols are used to plot measurements from different observatories, as indicated on the top of the Figure. }
    \end{center}
    \label{fig:fig3}
\end{figure*}

The above correction factors, which were of the order of a few percent,  were applied to the profiles displayed in Figure~1 and to the EW values presented in Figure~3. The weighted average of the resulting equivalent widths is $-$69.5 \AA\ with a standard deviation of 1.3 \AA. Again the small changes are most likely caused by clumps in the WR wind. 

Finally, we measured the skewness and kurtosis of the line and present the values as a function of time respectively in the third and bottom panel of Figure~3. As these quantities describe the shape of the line, the measurements are not affected by the blaze function problem mentioned above. The weighted mean for the skewness, $+$0.0374 is very small and the standard deviation is 0.0147.  The positive value of the former reflects the fact that during our observations, the line was slightly skewed to red wavelengths (the line flux on the blue side was higher than on the red side), as can be clearly seen in Figure ~1. This can possibly be caused by a faint colliding wind excess of this very wide binary.  Using the orbit and ephemeris of  \citet{2005MNRAS.360..141L} and \citet{2016MNRAS.461.4115R}, the separation between the stars is $\sim$440 R$_{\rm O} \sim$925 R$_{\rm WR}$ and the mid-point of our observing runs yields an orbital phase of about 0.3, which corresponds to the WR star behind the O star. As the shock cone is expected to be wrapped around the star with the smallest momentum flux, in this case the O star, this is exactly what is expected (a blue-shifted emission excess from the material flowing along the shock cone). Note however that this is a very wide system and that the detection of a colliding wind excess, even weak, is thus surprising.  In principle, such an asymmetry can be reproduced by spherical wind models. It arises  since rays that intersect the optically-thick part of the star produce only blue-shifted emission. However, the only way to determine which interpretation is correct would be to obtain a series of high signal-to-noise spectroscopic observations for a complete orbit and with an extremely high S/N, as in view of the large separation, the effect is expected to be small. Finally, the weighted mean value of the kurtosis, +2.144 with a standard deviation of 0.031, reflects the fact that the C{\sc iii}$\lambda$5696 profile for this star is rather flat-topped (a Gaussian profile would have a value of 3).

\section{The Nature of the WR137 Binary System}
\subsection{Implications of the Lack of Detection of Large-Amplitude Spectroscopic Variability}

One of the main reasons we originally included WR137 as a target in this observing campaign was because the level of polarization of the continuum light from this star detected by \citet{2000A&A...361..273H} was found to be variable with a peak-to-peak amplitude of $p\sim$0.3\% on a timescale of more than 5 years with no clear dependence on the orbital phase. The position angle on the other hand was found to be constant.  This  was interpreted as indicating the presence of a large-scale asymmetry in the wind of the WR star in this system. As suggested by those authors, we aimed to search for periodic spectroscopic variability associated with the presence of a potential CIR such as had been found in the two WR stars WR6 and WR134 also presenting line depolarization \citep[e.g.,][]{1997ApJ...482..470M, 2016MNRAS.460.3407A}.  Unfortunately, we detected only small scale variability over the four-month period of the observing campaign and therefore we were not able to confirm the presence of such a large-scale structure, at least in the time interval covered by our observations.  A CIR viewed pole-on would produce constant line profiles but also a constant polarization level and therefore such a geometry is incompatible with the spectroscopic and polarimetric data available for this star, although note that they were not obtained simultaneously.   There remains the possibility that we are indeed viewing a CIR pole-on if we attribute the mean polarization value to the CIR and the variability to clumps in the wind. However, this viewing geometry is highly unlikely.

We have searched for periodicities within the small-scale variability of the quantities describing the C{\sc iii}$\lambda$5696 spectral line (radial velocities, equivalent widths, skewness and kurtosis) using the Period04 package\footnote{https://www.univie.ac.at/tops/Period04/} but found no signifiant peaks.  Therefore, we are unable to confirm the period of 0.83 days found by \citet{2005MNRAS.360..141L}  in the absorption troughs of the C{{\sc iv}}$\lambda$5802/12 and He{\sc {i}}$\lambda$5876 spectral lines. 

\subsection{The O-star companion}
\subsubsection{The Spectrum}
	As noted in Section 1.1, the most up-to-date spectral type of WR137 from the spectral modelling of \citet{2016MNRAS.461.4115R} is WC7pd+O9V.  This was based on a spectral analysis of both binary components carried out with the Potsdam Wolf-Rayet \citep[PoWR; ][]{2004A&A...427..697H} NLTE model atmosphere code. The various parameters obtained for the WR star and the O-type companion can be found in Table\,\ref{WR137parameters}, which reproduces values from their Table~4. The best fitting model for the O star yields a projected rotational velocity of 220 km\,s$^{-1}$. Close inspection of our high-resolution  Keck  spectrum has revealed that not all absorption lines can be reproduced with such a rotation velocity.  In Figure 4, we present zooms of our mean Keck spectrum in the wavelength regions of the H$\alpha$, H$\beta$, H$\gamma$, He{\sc i}$\lambda$5876, He{\sc i}$\lambda$6678 and of the He{\sc i}$\lambda$4388, 4026 and 4471 lines. We plot all profiles in velocity space in order to facilitate comparison and use a different colour for each line (except for the He{\sc i}$\lambda$4026 and He{\sc i}$\lambda$4471 lines for which we use the same colour as for the He{\sc i}$\lambda$4388 line). In the panel on the right-hand side of the figure, we also show a montage of the profiles of these lines as well as that of the He{\sc i}$\lambda$4388 line,  on which we have superposed the model of \citet{2016MNRAS.461.4115R} for comparison.  For each transition, we used the same colour as in the individual panels.  We find that most helium absorption lines, here exemplified by the He{\sc i}$\lambda$4388, 4026 and 4471 lines, clearly have different characteristics from those of the hydrogen lines, as can be seen from this figure. While the He{\sc i} lines (illustrated by the $\lambda$4388 line in the plot in the right-hand panel), attributed to the O9 star, are fitted with a pure absorption component with the above-mentioned rotation velocity, the latter, if interpreted as in absorption, are much narrower and cannot be reproduced by such a wide absorption  (the dot-dashed vertical lines in the right-hand panel of Figure 4 indicate the limits of the observed and theoretical He{\sc i}$\lambda$4388 profiles). Instead, they can be viewed as two sharp emissions features centered on the rest wavelength of the lines and separated by several hundred km\,s$^{-1}$. It appears that two He{\sc i} lines,  He{\sc i}$\lambda$5876 and He{\sc i}$\lambda$6678, also exhibit a double-peak profile. Note that these two lines are the strongest in the $2^3P^0-n^3D$ and $2^1P^0-n^1D$ series of neutral helium, respectively. 

\begin{table}
\centering
\caption{Stellar and Wind Parameters for WR137 from \citet{2016MNRAS.461.4115R}.}
\label{WR137parameters}
\renewcommand{\arraystretch}{1.5} 
\begin{tabular}{lcc}
\hline
     & WR Star & O Star\\
\hline
Spectral Type & WC7pd & O9V\\
T$_*$ (kK) & 60$^{+5}_{-5}$  & 32$^{+2}_{-2}$\\
log g$_*$(cm s$^{-2}$)& -- & 4.0$^{+0.3}_{-0.3}$\\
log L (L$_{\odot}$)& 5.22$^{+0.05}_{-0.05}$& 4.75$^{+0.05}_{-0.05}$\\
log R$_t$ (R$_{\odot}$)& 0.7$^{+0.05}_{-0.05}$& -- \\
v$_{\infty}$& 1700$^{+100}_{-100}$& 1800$^{+100}_{-100}$\\
R$_*$(R$_{\odot}$)& 3.8$^{+1}_{-1}$& 7.7$^{+1}_{-1}$\\
log $\dot{\rm M}$& -4.65$^{+0.2}_{-0.2}$& -7.1$^{+1.0}_{-0.3}$\\
$v$\,$\sin$ i (km s$^{-1}$) & -- &220$^{+20}_{-20}$\\
M$_v$ (mag)& -4.18$^{+0.2}_{-0.2}$& -4.34$^{+0.2}_{-0.2}$\\
M$_H$ (mag)& -4.07$^{+0.2}_{-0.2}$& -3.41$^{+0.2}_{-0.2}$\\
E(B-V) (mag)& 0.74$^{+0.02}_{-0.02}$& \\
A$_v$ (mag)& 2.29$^{+0.06}_{-0.06}$& \\
\hline
\end{tabular}
\end{table}
	
\begin{figure*}
	\begin{center}\includegraphics[width=\textwidth]{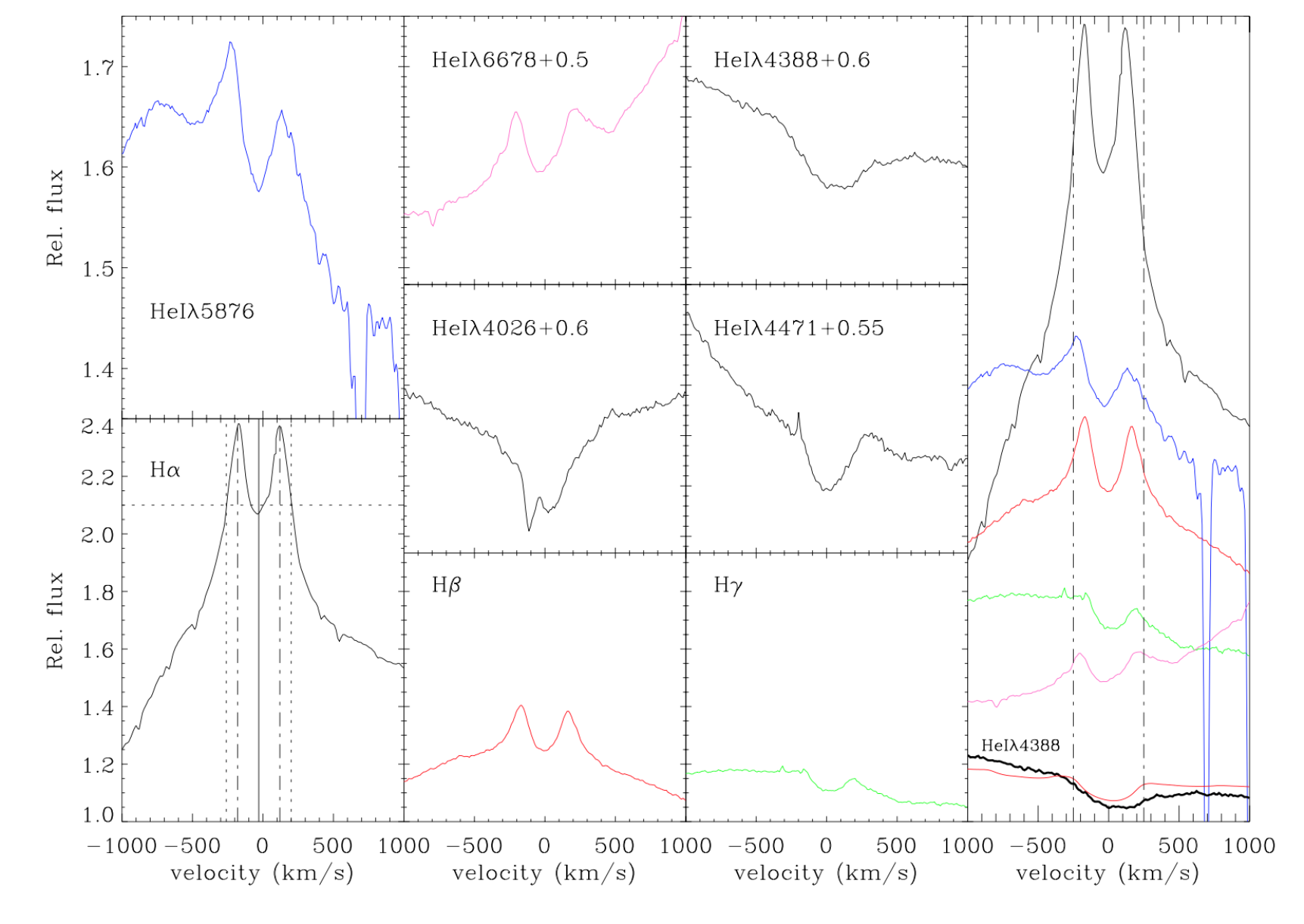}
    \caption{Zooms in velocity space of our Keck mean spectrum in the wavelength regions of the H$\alpha$, H$\beta$, H$\gamma$, He{\sc i}$\lambda$5876 and He{\sc i}$\lambda$6678 lines as well as He{\sc i} He{\sc i}$\lambda$4388, 4026 and 4471 lines (the sharp absorption superposed on the He{\sc i}$\lambda$4026 line is most likely interstellar in origin). We use a different colour for each line. The y-scale for the three hydrogen lines has been kept the same and the scale for all the He{\sc i} is different but also kept the same. In the right-hand panel we display a montage of all these lines.  We superimposed a PoWR model spectrum for a pure absorption profile (red) on the observed He{\sc i}$\lambda$4388 line (black).}
    \end{center}
    \label{fig:fig4}
\end{figure*}

\citet{1992ApJ...398..636U} had already reported the sharp double-peak emission profiles of the H$\alpha$, H$\beta$, H$\gamma$ and He{\sc  i}$\lambda$6678 lines that she attributed to a rotating thin, ring-like disk associated with the line-emitting region of the WR star. In that paper, variations in the intensity ratio of the blue and red components were reported on a very short timescale, going from larger to smaller than unity over a 4-day period.  \citet{2000A&A...361..273H} also report a double-peak profile for  the H$\alpha$ line. They hint to variability of this line by mentioning that {\em ``The H$\alpha$+He{\sc ii} complex at 6560 \AA\ shows a double-peak morphology  at  some  phases"}.  They attribute the double-peak profiles to an unusual morphology of the WR wind, 
likely associated with the continuum polarization they detect.   However, an examination of the plots presented by these authors reveals that the double-peak profiles are only visible in their higher resolution spectra (2-5 \AA ) obtained at the William Hershel Telescope but not in the lower resolution (8-10 \AA ) observations obtained at the Pine Bluff Observatory. This is to be expected because, as can be seen in Figure 2, the separation between the two peaks of the H$\alpha$ line is at most 8 \AA.   Therefore, we suggest that they were not able to identify the double-peak profile in their lower resolution data and therefore that there is no clear evidence that the profile was single-peaked at some epochs.

We carefully measured the blue-to-red intensity ratio of the double-peak profile for all the H$\alpha$ profiles in our dataset and find an average value 0.997$\pm$0.004 with a standard deviation of 0.051. Therefore, contrary to \citet{1992ApJ...398..636U}, we find no significant difference in the intensity  of both components and no clear variability over our 4-month observing period.

\subsubsection{The Nature of the Star}
We suggest that these double sharp-peaked profiles are unlikely to come from the WR star as all other emission lines are well reproduced by a wind with a terminal velocity of 1700 km\,s$^{-1}$ \citep{2016MNRAS.461.4115R}, whereas the separation between the peaks superposed on the hydrogen lines is at most 370 km\,s$^{-1}$ ($\sim$8 \AA ).  Furthermore, hydrogen is never found in the winds of WR stars of spectral type WC. Actually, \citet{1962ApJ...136...14U} had originally attributed the double-peak profiles of the H$\alpha$ and H$\beta$ lines to a Be star. Later, \cite{1981ApJ...246..145M} suggested that the origin of these apparently double-horned emission features was more likely absorption lines with the same origin as the other absorption lines in the spectrum.  Finally, \citet{1992ApJ...398..636U} later revised her interpretation and attributed the profiles to a large, thin, ring-like disk of cool plasma associated with the WR wind.  

We propose that these double-peak profiles are associated with the O9 companion and that it is in fact of spectral type O9e, i.e. the O companion  harbours a decretion disk. Although it is difficult to visually untangle the Oe spectrum from that of the WR star, we can attempt to characterize the double-peak line profiles in order to compare their characteristics to those from known  Be  disks.  Here we will assume that Oe stars are merely the extension of Be stars to earlier spectral types and follow the findings for Be stars that are much more numerous. 

\citet{1996A&A...308..170H} found a tight correlation between the total width of Fe{\sc ii} emission lines from the disk and the value of $v\, \sin i$ of the star. Unfortunately, these lines are not seen in our case, which is perhaps not surprising in view of the higher temperature of this Oe star compared to Be stars. For the H$\alpha$ line, correlations also exist but they are not as tight, most likely because this line is optically thick and therefore affected by non-kinematical effects. \citet{1989Ap&SS.161...61H} found a linear correlation between the Full Width at Half Maximum (FWHM) and the peak separation, $\Delta v_{\rm peak}$, of disk emission lines and $v\, \sin i$:
$${\rm FWHM(H}\alpha{\rm )=}1.4v\sin i + 50 {\rm km\, s^{-1}}$$
$${\Delta v_{\rm peak}\over v\, \sin i}=0.5.$$
The second equation applies for the majority of stars but many present a higher value of this ratio, although none is above 2.0.  

We measured both the FWHM and $\Delta v_{\rm peak}$ values for our mean H$\alpha$ profile that is superposed on a complex of WR emission lines (He{\sc ii}$\lambda$6560, C{\sc ii}$\lambda$6578 and C{\sc iv}$\lambda$6592). In Figure 4, in the bottom left panel where the H$\alpha$ profile is displayed, the dotted lines indicate our estimated half maximum level as well as the limits of the full width of the line at that level and the dot-dashed lines indicate the position of the two peaks. We obtain FWHM(H$\alpha$)=460 km\,s$^{-1}$ and $\Delta v_{\rm peak}$(H$\alpha$)=300 km\,s$^{-1}$.  The projected rotation velocity of the O companion was estimated from the spectral fitting of \citet{2016MNRAS.461.4115R} to be 220 km\,s$^{-1}$.  These values fit relatively well within the scatter of the above empirical relations found by \citet{1989Ap&SS.161...61H} (see their Figures 1 and 7). Our measured ratio between $\Delta v_{\rm peak}$(H$\alpha$) and $v\, \sin i$ is 1.4, well below the maximum empirical value of 2.0. Actually, \citet{1988A&A...189..147H} showed that this ratio also depends on the equivalent width of the profile. A value of 1.4 is more compatible with the relation found for weak profiles ($<$ 1.5 \AA ).  In view of the blend of the double-peak profile with the strong WR emission lines, it is difficult to determine if its equivalent width for WR137 is below this level.

\subsubsection{Polarization and Orientation of the Disk}
{Shell stars are Be stars characterized by two peaks separated by a strong and narrow absorption (or central reversal) that reaches below the flux coming from the stellar photosphere.} These double peaks are normally associated with an edge-on view, or nearly so,  of the decretion disk.  In our case, it is difficult to say if the central reversal between the peaks reaches below the regular flux from the WR wind particularly for the H$\alpha$ line. However, it is clear that the central absorption is quite strong and therefore compatible with the fact that we know we are viewing the WR137 binary system from quite a large angle, i.e. nearly edge-on \citep{1999ApJ...522..433M}. This would also indicate that the Oe-star axis is nearly aligned with that of the orbital plane. Shell stars also have among the largest peak separations among Be stars, which is compatible with our measurement described above \citep{2003PASP..115.1153P}. In this four-month campaign, we detected no variability in the values of the height of the blue-shifted and redshifted peaks, which as mentioned above have a ratio very close to 1.  This is the case for the majority of Be stars \citep{2003PASP..115.1153P}. 

\citet{2003PASP..115.1153P} also report that almost all Be stars are polarized in continuum light (up to 2\%) with constant polarization angles. This is compatible with the published polarization observations of WR137. Indeed, \citet{1998MNRAS.296.1072H} found an {\it intrinsic} level of polarization for WR137 of P$_c$=0.57$\pm$0.2 \%, which after taking into account that the O star contributes only 41\%\ of the total continuum flux, corresponds to an intrinsic polarization for the decretion disk of $\sim$1.4 \%.  Furthermore, \citet{2000A&A...361..273H} found a non-variable polarization angle of $\sim$17$^{\rm o}$, i.e. perpendicular to the plane of the binary orbit and therefore of the Oe disk if its axis is aligned to that of the binary, as hinted at from our observed double-peak profiles.  \citet{1997ApJ...477..926W} have shown that the observed levels of polarization and constant polarization angles can be reproduced by a thin keplerian or wind-compressed disk. However, the formation of this type of disk is inhibited because of non-radial radiative forces \citep{1996ApJ472115} .  Nevertheless, the current interpretation for Oe/Be stars, i.e. a decretion disk from a fast rotating star, is likely to lead to similar levels of continuum polarization. We therefore suggest that the intrinsic polarization of WR137 is not from large-scale structure in the WR wind, i.e. from a CIR or from a flattening of the WR wind, but from the Oe disk instead. This is compatible with the fact that our 4-month spectroscopic monitoring campaign has revealed no large-amplitude spectroscopic variability, which should have been found if a strong-enough CIR had been  present.  \citet{1989ApJ...347.1034R} also report a small level of continuum polarization variability that could, in this interpretation,  come from clumps in the WR wind. 

\subsection{Evolutionary Status}

The existence of a binary system consisting of a WR star and a Oe companion is intriguing. To our knowledge, no other such system is known in the Galaxy (e.g. the online catalogue of galactic WR stars maintained by P. Crowther: http://pacrowther.staff.shef.ac.uk/WRcat/) or in the Magelanic clouds
\citep{2018ApJ...863..181N, 2003PASP..115.1265M}.  Such a combination is obviously a rare occurence but is the fraction of Oe companions in WR+O systems compatible with the fraction of Oe stars among O stars? The fraction of Oe types among Galactic O stars is very low. Out of the 448 O stars in the GOSS catalogue \citep[Goss; ][]{2011ApJS..193...24S, 2014ApJS..211...10S}, there were only 13 Oe stars known or $\sim$3\%. \citet{2018ApJ...863...70L} found an additional six Oe stars in the Galaxy, which brings the fraction up slightly to $\sim$4\%. Note that the fraction of Oe stars in the SMC is much higher. \citet{2016ApJ...819...55G} found a fraction of $\sim$26\%.  The lower fraction in the Galaxy was explained by more angular momentum transport by stronger winds at higher metallicity suppressing the formation of the decretion disk.  The binary fraction of Galactic WR stars is thought to be $\sim$ 40\% \citep[e.g.][]{2007ARA&A..45..177C}. With about 430 WR stars known in the Galaxy, this corresponds to about 170 WR+O binaries. With only one Oe companion known, this is a fraction of 0.6\%. Of course, this is a strict minimum as not all spectral types of O companions are very well known.  We conclude that the fraction of Oe stars among Galactic O stars ($\sim$4\%) is not incompatible with the fraction of Oe type companions within WR+O binaries. This small fraction could be the result of the fact that decretion disks are thought to be short-lived around hot  luminous stars. \citet{2016MNRAS.458.2323K} have examined the ablation of disk material from the scattering of UV continuum photons from OB stars. For O stars, they find very short disk destruction times while for B stars it is quite a bit longer, which they use to explain the very small fraction of Oe stars among O stars when compared to the fraction of Be stars among B stars. According to their Figure 15, the lifetime of a decretion disk surrounding an O9 star is around 500ks or $\sim$6 days. Note however that in our case, we observe an extremely stable disk for the entire four-month observing period, which is incompatible with the above-mentioned ablation time. We also note that, as described in Section 4.2.1,  there is currently no convincing evidence in the literature indicating that the H$\alpha$ profile has ever shown a single peak profile.

\citet{2010MNRAS.405.2439O} obtained high angular resolution (0.07 $-$ 0.1$''$) infrared images for a sample of 40 B stars and 39 Be stars in order to search for companions and determine the binary fraction of both samples. These angular separations correspond to distances  of  20$-$1000 au and for such separations, binary components will only interact in highly eccentric systems.  For both samples, they find an identical binary fraction of $\sim$ 30\% and conclude that binarity cannot be responsible for the Be phenomenon in these wide binaries, although it does not exclude that it can happen in some cases.  On the other hand, the most likely descendants of {\em close} WR+Oe binary systems would be a High Mass X-Ray binary (HMXB). According to \citet{2006A&A...455.1165L}, about 60\%\ of HMXBs have a Be star as the non-compact component. Of course, these are close systems with the longest reported period being 262.0 days. But this high fraction strongly suggests that mass and angular momentum transfer in massive binaries is an important way of spinning up a main sequence star and leading to the Be/Oe phenomenon. The presence of an Oe star in a wide system such as WR137 could indicate that this can also occur in mass transfer in wide systems.  In this case, however, the SN explosion of the WR component is likely to unbind the system and not lead to a HMXB. 

The measured rotation rate of the O companion of $v\sin i$=220 km\, s$^{-1}$ is indeed faster than the bulk of O stars. \citet{2013A&A...560A..29R} measured the projected rotation velocities distribution of 216 presumably-single O-type stars in 30 Dor using the VLT-FLAMES Tarantula Survey and found a low-velocity peak at  $v \sin i\sim$ 80 km\, s$^{-1}$ with a high velocity tail extending to $v \sin i\sim$ 600 km\, s$^{-1}$.  \citet{2015A&A...580A..92R} measured the projected rotational velocity of 114 O-type spectroscopic binaries and found that the distribution for the O primaries presents a similar low velocity peak with shoulder at intermediate velocities (200<$v \sin i$<300 km\, s$^{-1}$). These results, which are consistent with what has been found from previous studies, strongly suggest that O stars in binaries are formed with the same spin distribution as single stars and that this distribution does not depend on the fact that these stars are in a binary or not.  Note that \citet{2017MNRAS.464.2066S} measured the rotation velocities of 8 O companions in WR+O binaries bringing the total number of measurements for such stars to ten. They find a much higher average velocity in He{\sc i} lines (348 km\,s$^{-1}$) than in He{\sc ii} lines (173 km\,s$^{-1}$), which they claim is evidence for strong gravity darkening as a consequence of fast rotation. However,  this result as been questioned as potentially been caused by an inadequate choice of the pseudo-continuum \citep{2018MNRAS.478.3133R}, shedding some doubt on the conclusion that O companions in WR+O systems have been spun-up by binary interaction.

Nevertheless,  the O-type companion in WR137 does have a spin rate compatible with having been spun up by a binary interaction and such a conclusion cannot be excluded. To explore the possibility that WR\,137 is the product of binary evolution, we utilise binary evolution tracks calculated with the  BPASS\footnote{bpass.auckland.ac.nz} (Binary Population and Spectral Synthesis) code  V2.0  \citep{2008MNRAS.384.1109E,2016MNRAS.462.3302E}. Each track is defined by a set of three parameters: the initial mass of the primary $M_{\rm i, 1}$, the initial 
period $P_\text{i}$, and the initial mass ratio $q_\text{i} = M_\text{i,2} / M_\text{i,1}$. The tracks  
were calculated at intervals of $0.2$ on $0.2 \le \log P\,[\text{d}] \le 4$,  $0.2$ on $0.1 \le q_\text{i} \le 0.9$, 
and at unequal intervals of $5-30\,M_\odot$ on $10 < M_\text{i,1} < 150\,M_\odot$. 

Following the approach described in detail in \citet{2016A&A...591A..22S, 2019A&A...627A.151S}, we find the best-fitting track through a $\chi^2$ minimisation algorithm that accounts for the observables of the system $(P, T_{*, WC}, T_{*, O}, \log L_{\rm WC}, \log L_{\rm O}, q=M_{\rm O}/M_{\rm WC})$  with their respective errors. The values reported by \citet{2016MNRAS.461.4115R} were obtained for a distance of 1.3\,kpc, which is significantly lower than the reported Gaia distance of 2.1\,kpc \citep{2018AJ....156...58B, 2020MNRAS.493.1512R}. To account for this, the luminosities of both components are revised upwards by roughly 0.4\,dex. Moreover, we allow for the temperature of the WC component to be arbitrarily high. The reason is twofold: first, WC stars are generally found in the degeneracy domain and their temperatures generally cannot be derived independently of M \citep[e.g.,][]{2003IAUS..212..198H}. Second, envelope inflation, which is believed to occur in WC stars and result in a lowering of the effective temperature \citep{2012A&A...538A..40G}, is not accounted for in the BPASS tracks. We do not consider the masses of $M_{\rm WR} = 5$ and $M_{\rm O} = 20\,M_\odot$ in our minimisation procedure, since these values depend on the calibration of the mass of the  O-component with its spectral type. As the spectra of both components were not yet disentangled, we consider the spectral type of the O-type secondary, and hence the absolute masses, as uncertain, especially in light of the scaled Gaia luminosities. The observables used for the minimisation procedure are given in Table\,\ref{tab:WR137_Pars}. 

\begin{table}\centering
    \caption{Inferred observables of WR\,137, adopted from \citet{2005MNRAS.360..141L} and \citet{2016MNRAS.461.4115R}, rescaled to the derived Gaia distance of 2.1\,kpc \citep{2020MNRAS.493.1512R}. The BPASS binary model has the parameters $M_{\rm i,1} = 60\,M_\odot$, $q_{\rm i} = 0.5$, and $\log P_{\rm i} = 3.2\,$[d], and an age of $4.1\,$Myr }
    \begin{tabular}{l|cc|cc} \hline \hline
       & \multicolumn{2}{c|}{Observables} & \multicolumn{2}{c}{BPASS model}\\ 
        Parameter & WC7pd & O9~V  & WC7pd & O9~V \\ 
        \hline
         distance\,[kpc] &  \multicolumn{2}{c|}{$2.1\pm 0.2$} & \multicolumn{2}{c}{-}\\        
         $\mathrm{P}_\mathrm{orb}$\,[d] & \multicolumn{2}{c|}{$4766 \pm 66$} & \multicolumn{2}{c}{$5030 $}\\
         $M_{\rm cur}$\,[$M_\odot$] &  $> 3.4$ & $> 15$ & 11 & 31 \\
         $q(\frac{M_{\rm O}}{M_{\rm WR}})$  &   \multicolumn{2}{c|}{$4.5 \pm 1.5$} & \multicolumn{2}{c}{2.8} \\
        $T_{\rm eff}$\,[kK] & > 60\,kK & $32\pm2$ & 120 &  35 \\
        $\log\,L\,[L_\odot]$  & $5.6\pm0.1$  & $5.2\pm0.1$  & 5.5 & 5.3 \\  
    \hline
    \end{tabular}
    \label{tab:WR137_Pars}
\end{table}

Out of the available tracks, we find that the binary track with $M_{\rm i, 1} = 60\,M_\odot, q_{\rm i} = 0.5$, and $\log P_{\rm i} = 3.2$\,[d] at an age of 4.1 Myr best represents the current parameters of the system out of the available tracks,  considering errors and grid spacing. The tracks are shown in Fig 5 and the reproduced BPASS values are shown in Table\,\ref{tab:WR137_Pars}. In the tracks, the primary reaches the RSG phase after roughly 3.6 Myr. At this stage, RLOF occurs and lasts roughly 5000\,yrs. During this phase, the primary loses roughly $12\,M_\odot$, of which $\approx 3\,M_\odot$ are transferred to the O-type companion.  By the end of the RLOF phase, the period grows by roughly 40\%. Further mass-loss since leads to a additional increase of the orbital period to $\approx 5000\,$d, which is close to the currently observed period. 

\begin{figure}
\begin{center}\includegraphics[width=\columnwidth]{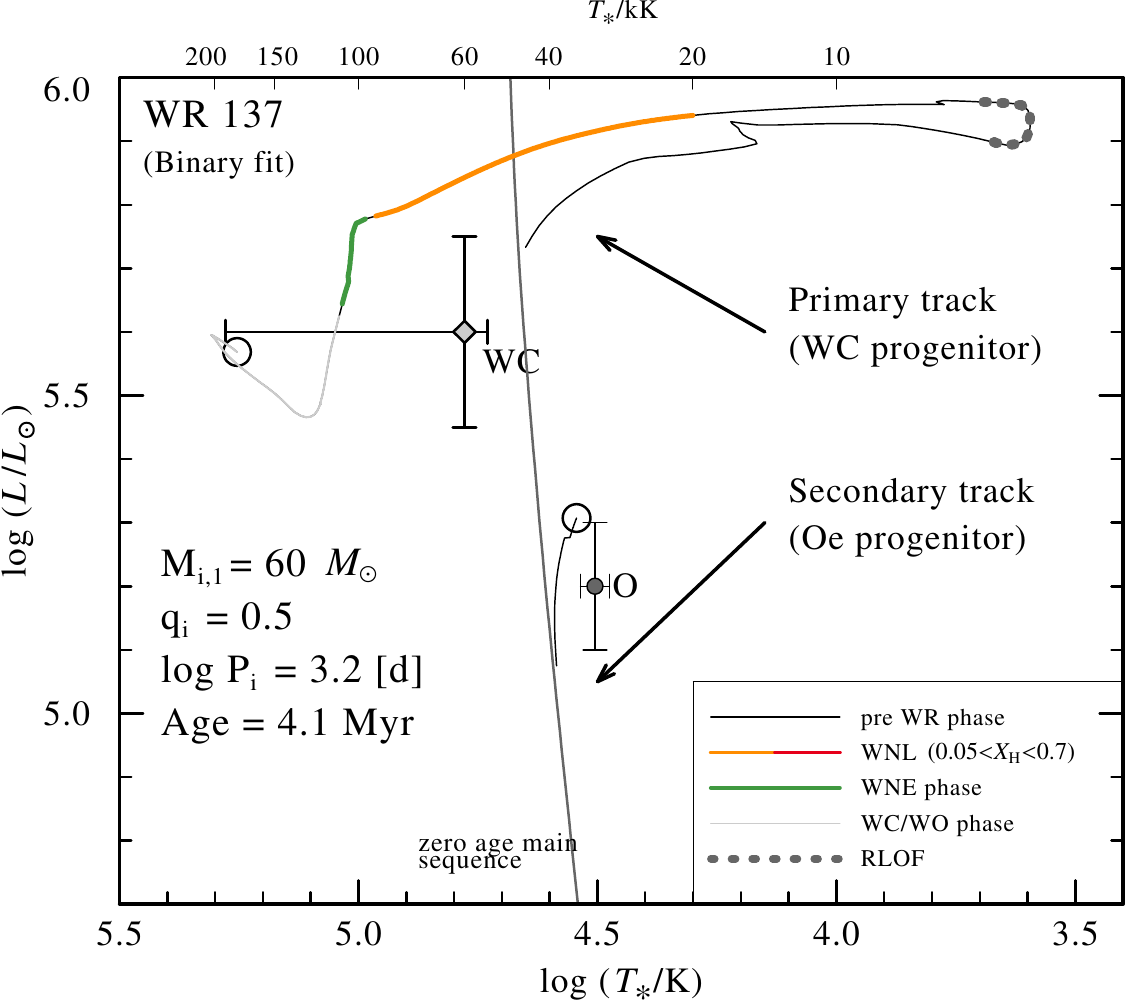}
    \caption{Best-fitting BPASS binary evolution tracks for WR\,137, corresponding to a binary with the initial parameters $M_{\rm i, 1} = 60\,M_\odot, 
q_{\rm i} = 0.5$, and $\log P = 3.2\,$[d]. Symbols with error bars mark the observed locations of the WC and Oe stars on the HRD, while the circles 
represent the corresponding best-fitting locations on the tracks at an age of 4.1\,Myr. }
    \end{center}
    \label{fig:fig5}
\end{figure}

We note that the mass ratio obtained in the BPASS model (2.8) is significantly lower than the reported value (4.5). Moreover, the absolute masses of $11\,M_\odot$ (WC) and $31\,M_\odot$ (O) are significantly larger than the masses obtained from the spectral type calibration discussed above ($5\,M_\odot$ and $20\,M_\odot$, respectively). However, considering the coarse grid spacing, the errors on the observables, and uncertainties related to mass-transfer efficiency and mass-loss, such discrepancies are acceptable. The BPASS solution should not be thought of as a tailored evolutionary path, but rather as a qualitative description of the binary evolution the system may have experienced. We further note that, while the primary reaches the RSG phase in the BPASS model, it is unclear whether a $60\,M_\odot$ progenitor would truly enter the RSG phase, considering the lack of observed RSGs at such initial masses \citep{1978ApJS...38..309H, 2018AJ....156...58B}. A further refinement of the parameters of the system (eg., through spectral disentangling) should help put further constraints on the question of past interaction between the two components.

Hence, it is conceivable that the Oe phenomenon originated in a post mass-transfer event, in which the WC progenitor transferred copious amounts of angular momentum to the secondary, making it rotate near-criticality.  It is not easy to explain how the Oe phenomenon should last for so long ($\approx 0.5\,$Myr) considering disk ablation due to the stellar radiation and wind \citep{2016MNRAS.458.2323K}. However, the mere fact that Oe disks do exist still needs to be explained in general. It is unclear at this point if their small numbers is compatible with their short expected lifetimes. 

\section{Conclusion}

In this paper, we have presented the results of a four-month spectroscopic observing campaign of the wide WR binary system WR137. Our observations were mainly concentrated on the C{\sc iii}$\lambda$5696 line but also included the region of the H$\alpha +$He{\sc ii} complex near 6560\AA\  for some of our observations. We found only small-scale variability in the profiles as well as in the integrated quantities describing the lines (radial velocity, equivalent width, skewness and kurtosis). These changes are most likely caused by clumps in the wind of the WR component of this system. 

The mean of the high-resolution, high signal-to noise spectra obtained at the Keck observatory allowed us to conclude that the H$\alpha$, H$\beta$, H$\gamma$ and the two strong He lines at $\lambda$5876 and $\lambda$6678 show a line profile that is different from that of most He absorption lines from the O companion, i.e. a double-peak emission. We therefore suggest that the companion harbours a decretion disk, which is compatible with its relatively large rotation velocity.

The presence of an Oe companion in WR137 is compatible with many of the observations of this star. The general characteristics of the double-peak emission lines we detect in the optical spectrum of this star are compatible with those of other Be/Oe stars. The intrinsic continuum polarization of this star, originally attributed to an asymmetry in the WR wind can now reasonably be attributed to a decretion disk around the O star instead. Our failure to detect large-scale spectroscopic variability potentially associated with CIRs in the wind of the WR stars during our four-month observing campaign is compatible with this interpretation. Furthermore, the interpretation of the X-Ray data of WR137 has been challenging.  \citet{2015MNRAS.447.2706Z} found that although the shape of the XMM-Newton spectrum of this star is compatible with models of colliding-wind binaries, the flux level is lower than predicted by two orders of magnitude, which would require a decrease in the mass-loss rate of the WR star by one order of magnitude. This author found that the spectrum can be reproduced with a two temperature optically thin plasma emission with kT$_1\sim$0.4 keV and kT$_2\sim$2.2 keV. Since Be stars are known to emit X-rays \citep[e.g.][]{2018A&A...619A.148N} perhaps an interpretation where one component is from the WR wind and the other from the Oe star should be considered and can possibly be acceptable to reproduce the X-ray observations. 

One of the most intriguing feature of the decretion disk around this O9e star, and perhaps around most Oe stars, is its stability. According to the current models, such a disk should be ablated in just a few days, while we have observed no variability, i.e. a high level of stability over a four-month period. There are surprisingly very few observations of the H$\alpha$ spectral region of this nevertheless well-studied star in the literature. It is possible that others have also been missed and that more such systems exist within the currently known WR+O population. Conversely, it is possible that O-companion decretion disks only form in yet-to-be-identified particular circumstances or conditions that are met in the case of the WR137 system, with ablation destroying the disk in most WR+O systems.  It is also unclear what the effects of the presence of the disk around the O star are on the dust formation by the wind collision in this system.  A careful spectroscopic study (perhaps at the next periastron passage in less than five years) together with hydrodynamic simulations of colliding winds could help constrain the geometry and strength of the two winds and determine to what extent they affect the wind-wind collision zone and thus the formation of the dust by this system.

\section*{Acknowledgements}
First, the professional astronomers authors of this paper are grateful to the amateur astronomers of the Teide team as well as to individuals observers, who invested personal time, money and enthusiasm in this project. NSL and AFJM wish to thank the National Sciences and Engineering Council of Canada (NSERC) for financial support. JK and BK are thankful for the support by grant 18-05665S (GA \v{C}R). The Astronomical Institute Ond\v{r}ejov is supported by project RVO:67985815 of the Academy of Sciences of the Czech Republic.

We acknowledge the help and support of the staff at the Observatoire du Mont M\'egantic (Qc) and at the Dominion Astronomical Observatory (BC) in Canada, the W. M. Keck Observatory in the US, the Ond\v{r}ejov Observatory (the Perek 2-m Telescope) in the Czech Republic and the Teide Observatory in Spain. The 0.82m IAC80 Telescope is operated on the island of Tenerife by the Instituto de Astrof{\'i}sica de Canarias (IAC) in the Spanish Observatorio del Teide. We acknowledge the support from the amateur spectroscopy groups, VdS and ARAS.

Finally, the authors wish to recognize and acknowledge the very significant cultural role and reverence that the summit of Maunakea has always had within the indigenous Hawaiian community.  We are most fortunate to have the opportunity to conduct observations from this mountain.
\\
\\
\noindent {\bf Data Availability}
\\
The data (fits files containing our spectra) underlying this article will be shared on reasonable request to the corresponding author. The measurements of the line profiles underlying this article are available in the article and in its online supplementary material.

\bibliographystyle{mnras}

\bibliography{St-LouisWR137} 

%







\bsp	
\label{lastpage}
\end{document}